# Detecting Concept Drift for the reliability prediction of Software Defects using Instance Interpretation

Zeynab Chitsazian[1] . Saeed Sedighian Kashi[1] . Amin Nikanjam[1]

**Abstract**
In the context of Just-In-Time Software Defect Prediction (JIT-SDP), Concept drift (CD) can occur due to changes in the software development process, the complexity of the software, or changes in user behavior that may affect the stability of the JIT-SDP model over time. Additionally, the challenge of class imbalance in JIT-SDP data poses a potential risk to the accuracy of CD detection methods if rebalancing is implemented. This issue has not been explored to the best of our knowledge. Furthermore, methods to check the stability of JIT-SDP models over time by considering labeled evaluation data have been proposed. However, it should be noted that future data labels may not always be available promptly. We aim to develop a reliable JIT-SDP model using CD point detection directly by identifying changes in the interpretation of unlabeled simplified and resampled data. To evaluate our approach, we first obtained baseline methods based on model performance monitoring to identify CD points on labeled data. We then compared the output of the proposed methods with baseline methods based on performance monitoring of threshold-dependent and threshold-independent criteria using well-known performance measures in CD detection methods, such as accuracy, MDR, MTD, MTFA, and MTR. We also utilize the Friedman statistical test to assess the effectiveness of our approach. As a result, our proposed methods show higher compatibility with baseline methods based on threshold-independent criteria when applied to rebalanced data, and with baseline methods based on threshold-dependent criteria when applied to simple data.

## 1 Introduction

Concept drift (CD) refers to the phenomenon where the statistical properties of a data stream change over time (Lu, Liu et al. 2018). In Just-In-Time Software Defect Prediction (JIT-SDP), CD can occur when the characteristics of software defects change over time due to changes in the software development process, changes in the software architecture, or changes in user behavior (McIntosh and Kamei 2018). CD can have a significant impact on the accuracy of JIT-SDP models. If CD is not detected and accounted for, the model may become less accurate over time, leading to false positives or false negatives (Dong, Lu et al. 2017). By detecting and addressing concept drift in JIT-SDP models, organizations can improve their ability to identify and address software defects before they become critical issues. This can lead to improved software quality and reduced costs associated

✉ Zeynab Chitsazian
z.chitsazian@email.kntu.ac.ir

Saeed Sedighian Kashi
sedighian@kntu.ac.ir

Amin Nikanjam
nikanjam@kntu.ac.ir

[1] Faculty of Computer Engineering, K.N. Toosi University of Technology, Tehran, Iran

with fixing defects after they have been released into production. Therefore, there is a need for concept drift detection (CDD) techniques that can identify when CD is occurring and adjust the model accordingly. These techniques can include statistical methods such as change point detection or machine learning algorithms that can adapt to changing data distributions.

Through studying more than 15 chronological datasets, Krishna and Menzies (Krishna and Menzies 2018) concluded that the accuracy of the SDP model changes over time (Kabir, Keung et al. 2019). This issue arises in dynamic environments and poses a challenge for data streaming learning. Overall, detecting concept drift is crucial for ensuring model stability and reliability over time, particularly in real-world scenarios where data distributions are likely to change frequently. A study of checking model stability has been performed for three and six months in (McIntosh and Kamei 2018). Although the research aims to examine changes in commit properties over time, it is not focused on determining the ideal length of time periods for CD and believe that this length varies for different projects. Therefore, in this work, we dynamically derived CD occurrence points for the JIT-SDP model. Conventional methods for CDD are typically divided into two categories: performance monitoring algorithms and distribution comparison algorithms. These detectors indirectly identify CD by either monitoring model performance or data distribution. In contrast, the method based on model interpretation directly relates CDD logic to changes in a model's argumentation. The reason for this argument is that, according to (Demšar and Bosnić 2018), the model explanation offers insight into the decision-making process of the model to make the results more transparent (Demšar and Bosnić 2018). They assert that their approach is superior to others; therefore, the methods proposed in this work are based on interpretation. However, model interpretation requires input data labels to detect CD. Therefore, we use instance interpretation, which does not require labeling of input data. We then compare our results with those obtained using the method based on model interpretation on defect data. In the study conducted by Lin et al. (Lin, Tantithamthavorn et al. 2021), it was demonstrated that there is low consistency between the outputs of general interpretation and local interpretation. As little as 55% of the most important metric of the studied projects' local JIT models were consistent with the global JIT model, indicating that the interpretation of global JIT models is insufficient to encompass the diversity of interpretations for all local JIT models. Therefore, in this paper, we presents a CDD method designed specifically for local JIT models. On the other hand, inaccurate prediction models may result from training SDP models on imbalanced datasets where the ratio of clean and defective modules is not equal (Chen, Fang et al. 2018). To mitigate the impact of class imbalance, previous studies have utilized class rebalancing techniques (Tantithamthavorn, Hassan et al. 2018). Since our work also deals with datasets that have class imbalance issue, we have employed class balancing techniques and evaluated their impact on the output of proposed CDD methods. However, these techniques may lead to bias in learned concepts and affect the interpretation of SDP models (Tantithamthavorn, Hassan et al. 2018). Therefore, in this study, we have examined the impact of these techniques on the performance of proposed CDD methods based on instance interpretation. Through a case study on 20 datasets of open-source projects, we aim to address the following research questions:

**RQ1** Which instance interpretation algorithm is the most effective for detecting CD among the proposed algorithms?

**RQ2** How do the proposed methods perform compared to baseline methods for detecting CD using monitoring criteria other than ER? (For example, measures that depend on class rebalancing such as Recall, Gmean, or measures that are independent such as AUC, ER) and what is the effect of class rebalancing on the proposed CDD methods?

**RQ3** To what extent is the method based on repeated predictions on unlabeled data effective? How does this method compare to interpretation-based methods?

In this work, we propose a CDD method based on instance interpretation. We evaluate the effect of data resampling on the results of our proposed method using four threshold-dependent measures namely Precision, Recall, Gmean, and F-measure, as well as three independent measures: Area Under the Curve (AUC), Matthews correlation coefficient (MCC), and Error Rate (ER) (Lin, Tantithamthavorn et al. 2021). In summary, our paper outlines the following contributions:

- We introduce a new approach for discovering CD points in evolving software defect predictions at the commit level.
- We compare of our proposed method, which is based on instance interpretation, with model interpretation to detect CD. We identify the most suitable instance interpretation algorithm from those proposed in the literature and examine the impact of class rebalancing on CDD results.
- We detect CD using monitoring measures other than ER, such as Recall, Gmean, and F-measure, which are dependent on class rebalance, as well as independent measures like AUC, ER, and MCC.
- We evaluate CD detection using repeated predictions.

In this paper, we organize our discussion as follows: Section 2 provides an overview of previous works related to this study, Sections 3 and 4 describe the proposed method and results respectively, while the concluding section presents our conclusions and future work.

## 2 Related works

In this section, we will briefly discuss previous related research in various fields, including concept drift, model interpretation, and rebalancing techniques. Furthermore, we have examined the application of these fields in SDP.

### 2.1 Concept drift detection

In the study by (Demšar and Bosnić 2018), a new CD detector was proposed that can be used independently of the learning method, making it compatible with any classifier. Their proposed method is based on changes in model interpretation and utilizes the Interactions-based Method for Explanation (IME) algorithm to describe predictors and classifiers. In our previous work (Chitsazian, Sedighian Kashi et al. 2023), we also used interpretation vectors with positive and negative effect of the IME algorithm at the model-level to discover CD on software defect data at the commit level and evaluated its performance. In this study, we build upon our previous work and achieve better results. CDD solutions are commonly used in streaming data analysis problems, which are complex to learn and analyze (Lu, Liu et al. 2018). To address this challenge, CDD methods have been developed in recent years (Yang, Al-Dahidi et al. 2019, Zenisek, Holzinger et al. 2019, Abbasi, Javed et al. 2021). Ekanayake et al. (Ekanayake, Tappolet et al. 2012) examined the impact of CD on the performance of the SDP model. The study by (Baena-Garcıa, del Campo-Ávila et al. 2006) used the DDM strategy (Gama, Medas et al. 2004) to detect CD in time-ordered defect datasets and employed the chi-square test with Yates continuity correction (Nishida and Yamauchi 2007) to evaluate it. They also analyzed the presence of CD in SDP datasets. Kabir et al. utilized the DDM method at the inter-version level for CD detection (Kabir, Keung et al. 2020).

As previously mentioned, CD can cause changes in interpretation and stability of model performance, where stability refers to consistent performance over time, even after a year (Bangash, Sahar et al. 2020). Bangash et al.'s study aimed to examine the stability of performance measures of inter-project SDP models such as G-Measure, MCC, AUC, and f-score while considering version order accuracy since models should not train on future knowledge to predict past defects.

## 2.2 Model interpretation

Model interpretation refers to the process of understanding and explaining the results of a statistical or machine learning model. This involves analyzing the model's outputs to gain insights into how it makes predictions or classifications. Interpretation can involve examining the coefficients or weights assigned to different features in a linear regression or logistic regression model, analyzing the decision boundaries or feature importance scores in a tree-based model like random forests or gradient boosting, or using techniques like partial dependence plots and SHAP values to understand how individual features contribute to predictions in more complex models like neural networks (Massey 2011, Chatzimparmpas, Martins et al. 2020, Liang, Li et al. 2021). The goal of model interpretation is to gain a deeper understanding of how a model works and why it makes certain predictions. This can help improve its accuracy, identify potential biases or errors, and build trust with stakeholders who may be relying on its results. Recent literature has proposed model-agnostic techniques for interpreting the prediction of black-box AI/ML algorithms by detecting the contribution of each metric in predicting the class label of an instance (Jiarpakdee, Tantithamthavorn et al. 2020). Jiarpakdee et al. have experimentally evaluated three model-agnostic techniques: two advanced techniques of Local Interpretability Model (LIME) (Ribeiro, Singh et al. 2016) and BreakDown technique (Gosiewska and Biecek 2019) in their study. LIME utilizes a local regression model around the instance to calculate the contribution of each metric in predicting the instance class. On the other hand, BreakDown breaks down the prediction of the instance label into parts for interpretation, displaying the contribution of each metrics separately in predicting the target class (Jiarpakdee, Tantithamthavorn et al. 2020). In a study by (Demšar and Bosnić 2018), a method based on model interpretation was used to discover CD, claiming that it works better than data performance monitoring methods such as ER. The IME method is used for model interpretation to obtain CD through distance analysis of model interpretation vectors. The study is the only paper that used interpretation to detect CD. The IME algorithm returns three types of interpretation vectors at the model level: positive effect size that features have on the target class label, negative effect size, and average effect size. In the research, the last case was used to detect CD, but we discover that the other two types generally give better results. Another method we use to discover drift is detection by changing the distribution of normalized and not interpreted data vectors. We compare this with other methods using raw data distribution change methodology.

## 2.3 The effect of rebalancing techniques on performance and interpretation

The performance and interpretation of an SDP model heavily depend on the dataset used. SDP models trained on imbalanced datasets are highly prone to producing incorrect predictions. The proportion of defective and clean components differs in imbalanced datasets, resulting in poor identification of data containing defects (He and Garcia 2009). Previous studies have utilized class rebalancing techniques to reduce the effect of class imbalance in the data by generating equal numbers of defective and clean components before building defect prediction models. Many studies have shown improved performance when using rebalancing techniques, such as Chawla (Chawla 2003) and Seiffert (Seiffert, Khoshgoftaar et al. 2009), who demonstrated that AUC can improve by up to 40% by applying rebalancing techniques. However, Riquelme et al. (Riquelme, Ruiz et al. 2008) concluded that rebalancing techniques do not significantly affect the performance of defect prediction models, leading to conflicting conclusions. Moreover, Riquelme et al. point out that using class rebalancing techniques may lead to CD because statistical distributions of training and test datasets differ. Therefore, class rebalancing techniques may affect the interpretation of defect prediction models. Tantithamthavorn et al. (Tantithamthavorn, Hassan et al. 2018) concluded a study on the impact of four commonly used rebalancing techniques, namely SMOTE, ROSE, over-sampling, and under-sampling, on the performance and interpretation of SDP models trained with seven well-known classifiers. We then evaluated the performance of these models using seven well-known measures: three threshold-independent measures (e.g., AUC) and four threshold-dependent measures (e.g., Precision, Recall). They found that the impact of rebalancing techniques on SDP models' performance over time varies depending on the classification algorithm and evaluation metrics used. However, AUC remains unaffected by these techniques and should be considered a standard measure for evaluating SDP. Unfortunately, conventional machine learning models fail to accurately classify and predict imbalanced datasets resulting in low accuracy for minority samples.

## 2.4 Verification latency in SDP

On the other hand, training examples generated in online problems often face the challenge of verification latency (VL). For instance, a software change is labeled as either clean or defective when a defect is found or after a specifically defined time has passed (whichever comes first). If no defects are found during this waiting period, it is assumed that the desired change is clean and labeled as a clean training class (Tabassum, Minku et al. 2022). Therefore, any training samples that are not identified as defective considering the VL are available for training the model after a specific time (Cabral, Minku et al. 2019). According to Cabral, Minku et al., "Defect discovery delays ranged from 1 day to over 11 years, and medians were typically close to or lower than 90 days. A waiting time of 90 days can be considered to provide a good trade-off between correct labeling and CD". For this reason, we consider distance as VL between training and test data. Based on our investigations, we have found that the time interval between each 100 samples in our dataset is approximately 90 days or more.

## 2.5 Related evaluation methods

In this section, we will discuss the evaluation methods and criteria presented in previous related works. Gama (Gama 2010) identified ER change points using the PH statistical test as a CDD method, which Demšar used as a baseline method (Demšar and Bosnić 2018). Similarly, we have also used this baseline method to evaluate our work. Figure 1 displays the error rate monitor over time, where CD points can be recognized. The central figure represents the cumulative ER graph, while the right figure shows the application of the PH method. Ekanayake (Ekanayake, Tappolet et al. 2009) also utilized AUC prediction for change detection. McIntosh et al. (McIntosh and Kamei 2018) investigated the effect of fluctuating features values on the performance and interpretation of the JIT model by monitoring the efficiency of AUC and Brier Score performance measures. Bangash et al. (Bangash, Sahar et al. 2020) evaluated the stability of cross-project defect prediction models by measuring their performance estimates in different periods, considering G-measure, MCC, AUC, and F-Score as performance measures. Gangwar et al. (Gangwar, Kumar et al. 2021) proposed a paired learner-based drift detection method in SDP that examines a subset of data at the same time by two learners (stable and reactive) to detect drift by exploring the dissimilarity of their predictions using AUC as a measure of model instability. Kabir et al. (Kabir, Keung et al. 2021) assumed that previous software releases are labeled (clean or defective) to train inter-release DP models (IRDP), evaluating their strength against CD using AUC, Recall, and pf performance stability measures. The conventional CDD method monitors error performance through Error-Rate (ER) (Demšar and Bosnić 2018), which can be calculated using Equation 1.

$$\text{Error Rate} = 1-\text{Accuracy} \qquad (1)$$

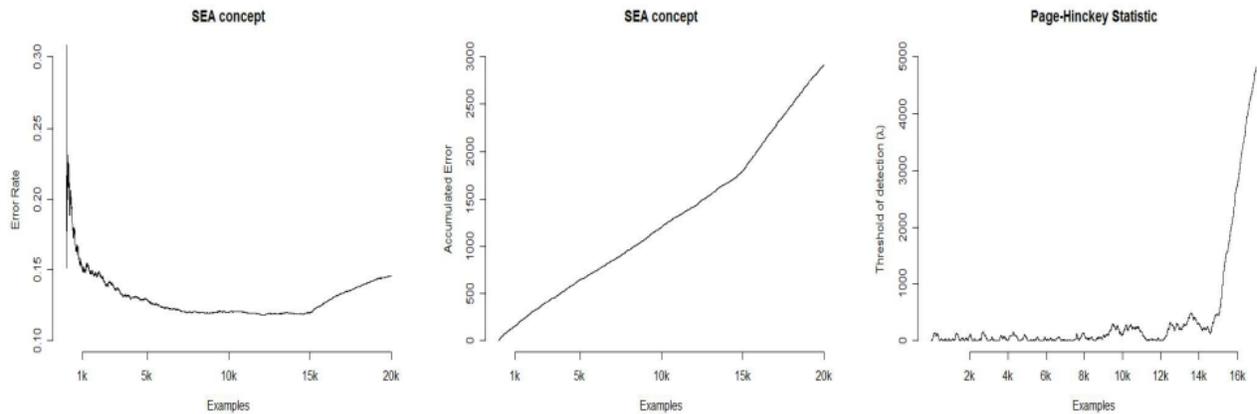

**Fig. 1** The curve on the left represents the ER graph over time. The central graph shows the cumulative ER, while the right graph illustrates the application of the PH method. The slope changes at the drift event location, but it is more noticeable in the right graph. (Gama 2010)

## 3 Case Study Design

In this section, we describe the design of our proposed techniques and evaluation methods. Following that, we will explain our case study in order to address the research questions.

### 3.1 Dataset

In this study, we present the first empirical investigation into discovering of CD points in unlabeled data for the evolving JIT-SDP problem. Our study addresses the RQs by utilizing 20 GitHub repositories that were previously used in other studies (Lin, Tantithamthavorn et al. 2021). Lin et al. employed this dataset to derive the most important features in the SDP-JIT problem and investigated the consistency and stability of prediction performance across different datasets and models. Additionally, this dataset has two key characteristics: public availability and long-term development, making it an ideal resource for our study. Table 1 provides a summary of this dataset, which is contained three categories. Commit level metrics are concisely described in Table 2.

### 3.2 Data preprocessing

The datasets used in this study contain commit-level metrics that are heavily skewed and vary in scale, with some metrics showing high correlation. To address this problem, we followed a similar approach to the work of Lin et al. (Lin, Tantithamthavorn et al. 2021). Specifically, we centered and scaled the commit-level metrics using the scale function in R, except for the "FIX" metric which is Boolean. We also used Spearman correlation to mitigate collinearity and manually removed highly correlated metrics with a Spearman correlation coefficient greater than 0.7. Additionally, we normalized the entropy $\log_2^n$ by the maximum entropy $\log 2\ n$ to consider differences in the number of files across changes, as suggested by Hassan (Hassan 2009). We used the Synthetic Minority Oversampling Technique (SMOTE) proposed by Chawla et al. (Chawla, Bowyer et al. 2002) for class rebalancing to overcome the limitations of traditional oversampling and undersampling techniques. This technique generates artificial data based on feature space, rather than data space, and combines synthetic oversampling of minority defective modules with undersampling of majority clean modules. The SMOTE technique has an adjustable parameter that needs to be specified. However, Tantithamthavorn et al. demonstrated that changing the parameter of the SMOTE technique does not affect their findings (Tantithamthavorn, Hassan et al. 2018). We applied the SMOTE technique using the R package implementation of the SMOTE function (Torgo and Torgo 2013).

Table 1 Summary of software projects from the studied dataset (The percentage of defect commits, compared to clean ones, is shown in parentheses) (Lin, Tantithamthavorn et al. 2021)

| Project name | Date of first commit | Lines of code | # of changes |
|---|---|---|---|
| accumulo | Oct 4, 2011 | 600,191 | 9,175 (21%) |
| angular | Jan 5, 2010 | 249,520 | 8,720 (25%) |
| brackets | Dec 7, 2011 | 379,446 | 17,624 (24%) |
| bugzilla | Aug 26, 1998 | 78,448 | 9,795 (37%) |
| camel | Mar 19, 2007 | 1,310,869 | 31,369 (21%) |
| cinder | May 3, 2012 | 434,324 | 14,855 (23%) |
| django | Jul 13, 2005 | 468,100 | 25,453 (42%) |
| fastjson | Jul 31, 2011 | 169,410 | 2,684 (26%) |
| gephi | Mar 2, 2009 | 129,259 | 4,599 (37%) |
| hibernate-orm | Jun 29, 2007 | 711,086 | 8,429 (32%) |
| hibernate-search | Aug 15, 2007 | 174,475 | 6,022 (35%) |
| imglib2 | Nov 2, 2009 | 45,935 | 4,891 (29%) |
| jetty | Mar 16, 2009 | 519,265 | 15,197 (29%) |
| kylin | May 13, 2014 | 214,983 | 7,112 (25%) |
| log4j | Nov 16, 2000 | 37,419 | 3,275 (46%) |
| nova | May 27, 2010 | 430,404 | 49,913 (26%) |
| osquery | Jul 30, 2014 | 91,133 | 4,190 (23%) |
| postgres | Jul 9, 1996 | 1,277,645 | 44,276 (33%) |
| tomcat | Mar 27, 2006 | 400,869 | 19,213 (28%) |
| wordpress | Apr 1, 2003 | 390,034 | 37,937 (47%) |

Table 2 Summary of commit level metrics (Lin, Tantithamthavorn et al. 2021)

| Category | Name | Description |
|---|---|---|
| Diffusion | NS | Number of modified subsystems |
| | ND | Number of modified directories |
| | NF | Number of modified files |
| | Entropy | Distribution of modified code across each file |
| Size | LA | Lines of code added |
| | LD | Lines of code deleted |
| | LT | Lines of code in a file before the commit |
| Purpose | FIX | Whether or not the commit is a defect fix |

## 3.3 Constructing CDD methods

In this section, we clarify the construction of proposed and baseline CDD methods, thereby answering the RQs.

### 3.3.1 Discovery of CD points using instance interpretation based method (RQ1, RQ2)

A. Presentation of proposed methods

The purpose of this section is to provide a detailed description of the proposed method. In the study by Demšar and Bosnić, a method based on model interpretation was used to discover CD (Demšar and Bosnić 2018). According to their study, this method outperformed performance monitoring methods such as ER. However, this method presents two challenges. Firstly, it requires data labeling which may not be available for future data. Secondly, as mentioned in (Tantithamthavorn, Hassan et al. 2018) study, the use of resampling methods alters model interpretation. As a result, we utilized instance interpretation instead of model interpretation in this work and examined its effectiveness in detecting CD compared to the method that employs model interpretation (Chitsazian, Sedighian Kashi et al. 2023). Additionally, instance-level interpretation does not require labeling input data, making it more practical in real-world scenarios. In the proposed method, we assume that new data labels are unavailable; thus, we use instance interpretation. Whenever there is a change in instance interpretation, it indicates the occurrence of CD. We investigate the points in of the instance interpretation vector where changes occurred and evaluate whether they conform to the points of change in the model's performance using statistical methods. In (Jiarpakdee, Tantithamthavorn et al. 2020) study on finding a stable instance interpretation method in the SDP problem, it was asserted that instance interpretation with the BreakDown algorithm has better stability than the famous LIME algorithm. Therefore, we compare the effect of these two interpretation methods, BreakDown and IME, to detect CD. If we consider changes in ER as CD, this work shows that IME performs better than BreakDown in detecting CD points because its output is more consistent with the CD points discovered from the method based on ER monitoring. To demonstrate this, we evaluated the obtained CD points according to other important metrics such as AUC.

In (Lin, Tantithamthavorn et al. 2021) study, ANOVA Type-II variance analysis was used to extract the most important features. The aim was to compare the interpretation outputs of local and general, as well as combined project-oriented commit data, in order to determine their overlap. Additionally, multivariate analysis of variance (MANOVA) was employed on both raw and interpreted data to examine the statistical differences in interpretation vectors across various test data categories. For instance, if a dataset contains 9000 commits and we consider the test dataset as hundreds, we would have 90 test categories. Our method involves the following steps:

1- Data preprocessing, such as the removal of correlated features and rebalancing the class distribution.
2- Dividing the test data into groups of 100 for each of the 20 datasets as illustrated in Figure 2.

| 1-100 | 101-200 | 201-300 | | | | | | | 8901-9000 |
|-------|---------|---------|--|--|--|--|--|--|-----------|
| Group1 | Group2 | Group3 | | | | | | | Group9 |

**Fig. 2** In the first group, commits numbered 1 to 100 are arranged in chronological order until the end (this is done separately for all datasets). Each commit includes several properties, namely fix, nf, entropy, lt, la, and ld.

3- Deriving instance interpretation vectors for each group of data.
4- Extracting interpretation vectors that demonstrate statistical differences between different categories using the MANOVA method (these differences are shown in Table 3). A p-value less than 0.05 indicates a difference in distribution between two groups. Each column in Table 3 represents a feature, and each row corresponds to a test data category. The value in each cell shows the number of interpretation vectors from that group which have statistical differences from previous groups. We calculated the sum for each row in Table 3 and included it as the last column of the table (sum column).

**Table 3** The statistical distribution of the "FIX" feature vector in the 29th group significantly differs from that of the two previous groups.

|    | FIX | NF | ENTROPY | LT | CHURN | SUM |
|----|-----|----|---------|----|----|-----|
| 27 | 0 | 0 | 0 | 0 | 0 | 0 |
| 28 | 0 | 0 | 0 | 1 | 0 | 1 |
| 29 | 2 | 0 | 2 | 3 | 0 | 7 |
| 30 | 0 | 0 | 0 | 0 | 0 | 0 |
| 31 | 0 | 0 | 0 | 0 | 0 | 0 |
| 32 | 0 | 2 | 3 | 6 | 0 | 11 |
| 33 | 5 | 1 | 3 | 6 | 0 | 15 |
| 34 | 6 | 6 | 11 | 31 | 0 | 54 |
| 35 | 6 | 1 | 19 | 27 | 0 | 53 |

5- In the "sum" column, a value that significantly differs from the previous rows indicates the occurrence of CD. This suggests that the characteristics of the category have more statistical differences from earlier categories. Wherever consecutive rows in the "sum" column have more numbers than before, it indicates that this difference is non-random and significant. We locate these phase difference points using the PH algorithm. If only one row has a considerable difference, it may be a coincidence. For example, in Table 3, group 29 values in the "sum" column are increasing, representing a concept change because previous values are mainly 0 or 1.

6- We used Table 3 to calculate output for all 20 datasets using different interpretation methods and raw data.

7- We calculated the CDD evaluation measures for all three methods, which included two instance interpretation methods and normalized raw data distribution, on 20 datasets.

8- In order to compare the three methods, we used the Friedman test with k = 3 algorithms and n = 20 test groups (number of datasets).

9- We collected output measures from the Random Forest model, such as ER, AUC, Gmean, Precision, F-measure, Recall, and MCC for each category. These measures are commonly used in research to assess the stability of the SDP model.

10- We investigated at which points CD has occurred for each of the 20 datasets by monitoring each of these criteria using the PH statistical test.

For example, Figure 3 displays the output of the proposed method on the WordPress dataset. The ER PH method identifies points of CD occurrences, as presented in Table 4 . Figure 4 illustrates the segmented diagram of Figure 3, highlighting changes of concept in these areas that correspond to CDs obtained from the baseline method (Table 4). Unlike existing CDD methods, our approach controls false positive detection rates and enhances CDD algorithm performance by adjusting parameters. Suppose we have a stream containing 500 observations triggering a standard CDD method. Additionally, let us assume that CD points are identified at five different points. We want to determine whether these are false positives or actual CD points detected by the detector. If one CD is generated every 100 observations, all detected points are likely false positives. However, if a detector identifies CD points at different intervals, we can conclude with some degree of confidence that the change points are likely genuine (Ross, Adams et al. 2012). In this study, when using default parameters for the PH algorithm as described in (Demšar and Bosnić 2018), we encountered a challenge where CD occurs in all local models and on all datasets at almost the same points. The default parameters for PH were a warning threshold (corresponding to the allowable false alarm rate) $\beta = 0.01$, sensitivity (acceptable error rate) $\delta = 0.001$, and fading factor (update weight for historical values in the PH statistic) γ = 0.999 (Demšar and Bosnić 2018). This challenge exists when using this value or smaller values for the false alarm parameter. To address this issue, we increased the false alarm value to β = 0.1 and recommend checking the CD using other values of this parameter.

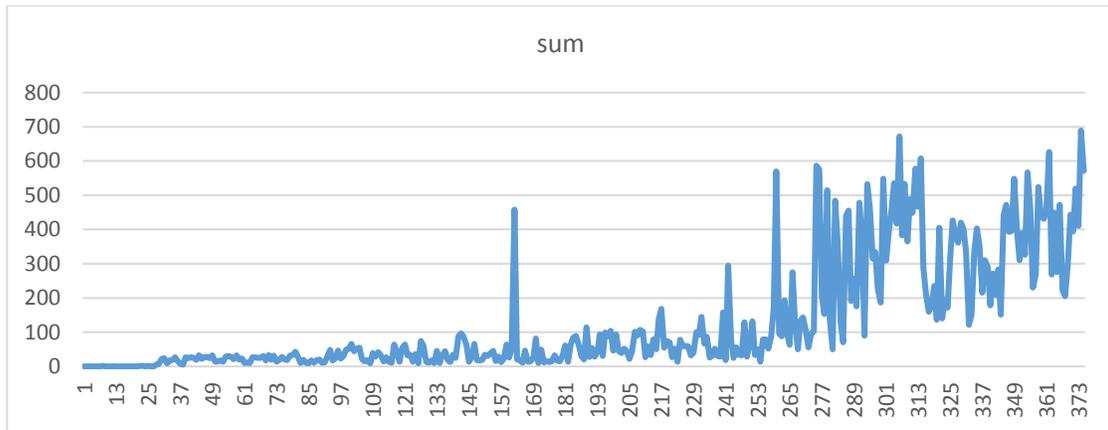

**Fig. 3** The sum column of Table 3 obtained from the WordPress dataset

**Table 4** Discovered CD points of WordPress dataset using ER PH method

| 24 | 53 | 82 | 111 | 142 | 174 | 203 | 232 | 261 | 290 | 325 | 357 |
|---|---|---|---|---|---|---|---|---|---|---|---|

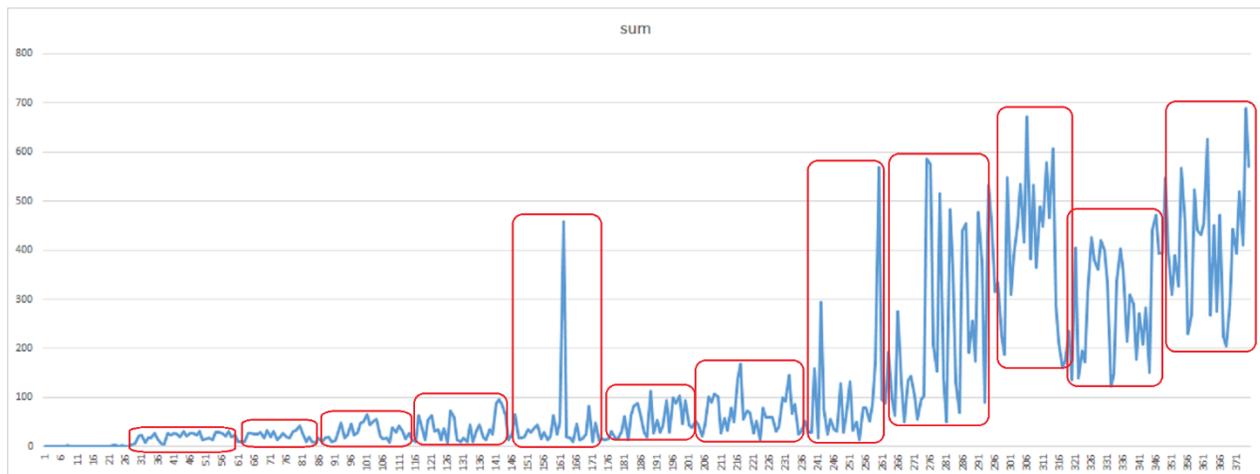

**Fig. 4** Figure 3 segmentation

B. Evaluation of the resampling-based methods

We used resampling techniques to address the imbalanced class distribution in the software defect datasets. However, Tantithamthavorn, Hassan et al. have stated that resampling techniques can alter data distribution and may lead to incorrect recognition of CD points (Tantithamthavorn, Hassan et al. 2018). To address this issue, we employed an interpretation-oriented CDD approach. However, since resampling can change the most important features of the prediction model, the CDD method based on model interpretation may inaccurately locate CD points. Therefore, we used instance interpretation to solve this problem. Finally, we evaluated our proposed CDD method on resampled data. The Friedman test revealed that our methods did not exhibit significant differences as in

(Demšar and Bosnić 2018) findings but yielded promising results. In Section 4, we present exciting results from the ranking of method outputs on 20 datasets.

C. Constructing Baseline Methods (RQ2)

In this section, we will discuss the baseline evaluation methods and criteria used to compare our work. In the methods, CD points are detected on the labeled data. Specifically, we measure model performance by monitoring the error rates over time using various criteria such as ER, AUC, and MCC. To detect CD, we monitor the performance of the error rate on different performance measures of the Random Forest machine learning algorithm using the Page-Hinkley test (PH) algorithm as described in (Demšar and Bosnić 2018) study. Whenever there is a change in the model performance, it indicates a CD point. We also monitor the performance of the Error Rate using PH algorithm, abbreviated as ER PH (Gama 2010). This algorithm can obtain CD points by monitoring the Error Rate performance over time. To evaluate model stability, different papers use varying performance measures of machine learning algorithms over time. In this study, we derived CD points using error monitoring of three threshold-independent performance measures (ER, AUC, and MCC) and four threshold-dependent measures (Tantithamthavorn, Hassan et al. 2018), namely Recall, Kappa, Precision, and Gmean, which we abbreviated as ER PH, AUC-Er PH, Recall-Er PH, Kappa-Er PH, Precision-Er PH, and Gmean-Er PH. These cases are baseline methods that we compare with the output of our proposed methods. For example, Equation 2 shows how AUC error or AUC-Er is calculated.

$$\text{AUC-ER} = 1 - \text{AUC} \qquad (2)$$

So far in this section, we have introduced metrics that can be monitored over time to locate CD. These are model performance metrics such as AUC, Recall, etc. Prior to this, only the ER has been utilized for detecting CD. We trained the model using Random Forest. Below, we provide appropriate evaluation metrics for CD identification to compare our proposed methods with baseline methods (Bifet, Read et al. 2013, Demšar and Bosnić 2018):

- CDD_Accuracy: The accuracy of the proposed method for detecting CD points.
- Missed Detection Rate (MDR): The possibility of not receiving a warning when there was a CD by calculating the ratio of unknown CDs to complete CDs (a proper detector will have an MDR equal to small or zero). MDR is obtained using Equation 3.

$$\text{MDR} = 1 - \text{CDD\_Accuracy} \qquad (3)$$

- Mean Time to Detection (MTD): The average delay between the detection of a CD point and its original CD occurrence location. A suitable detector should have a small MTD value.
- Mean Time between False Alarms (MTFA): This measure determines the duration of a false alarm in the absence of any CD occurrence. A suitable detector should have a high MTFA value.
- Mean Time Ratio (MTR): This is a measure of the trade-off between sensitivity and robustness, which can be obtained through Equation 4:

$$\text{MTR} = \frac{MTFA}{MTD} \times (1 - \text{MDR}) \qquad (4)$$

**3.2 The methodology of changing the distribution of frequent forecasts (RQ3)**

In literature, MCC and Brier Score measures are calculated using distance between the predicted and actual class probability by considering the test data label. Whenever this distance significantly increases, it is assumed that a CD has occurred (Gangwar, Kumar et al. 2021). However, we propose a new method that identifies CD by checking the distance between repeated predictions without considering the class label of the test data. Our approach involves investigating the distribution of frequent predictions. For each test set, we derived ten predictions and analyzed the distribution of categories that differ from previous categories using multi-variate one-way ANOVA of repeated predictions. We perform this evaluation in the following cases:

- Both with and without knowing the class label

- Both simplified and rebalanced data

# 4 Case study results

This section presents the results of proposed methods for each research question and compares them to those of the baseline methods.

## 4.1 Comparison of interpretation-based methods (RQ1)

For the reasons outlined in Section 3.1, we utilized two interpretation algorithms, IME and Breakdown (BD), to address RQ1. We compared the results of methods based on these two interpretation algorithms with each other and with the case where no interpretation was used. We named these three methods IME_base, BD_base, and raw_base, respectively. Additionally, we compared these three methods with the interpretation-based method performed on resampled data (rIME_base and rBD_base). Subsequently, we evaluated the degree of matching between the output of the proposed methods and each baseline method using CDD evaluation measures mentioned in Section 3.2 (cdAccuracy, MTFA, MTD, and MTR). Table 5 presents the evaluation results of the MTR criterion related to the output of the proposed CDD methods compared to the method based on ER monitoring. The last row provides the results of Friedman test. The lower rank number indicates better performance for that method. The method based on IME interpretation outperforms BD and raw_base (Table 5). Even in the interpretation-based method on rebalanced data, IME performed better. When we evaluated the proposed methods using the MTR measure compared to baseline methods based on AUC, Recall, and other criteria, we found that they gave the same result in all cases (Table 6). Friedman test results on other CDD measures are presented in Table 6 through Table 9.

Figure 5 displays a radar diagram that pertains to Table 6 through Table 9. This type of graph is used to compare the consistency of the output of the proposed interpretation-based methods with that of the baseline methods, which were influenced by four well-known performance criteria of CDD methods: MTR, cdAccuracy, MTD, and MTFA. Since the graph is associated with the Friedman test output, the closer it is to the center, the better it performs. As depicted in Figure 5, it is evident that the orange graph (i.e., IME-based method) consistently falls within the gray graph (i.e., BD-based). Consequently, it is apparent that the IME-based method consistently outperforms BD-based even in resampled mode (as indicated by the yellow graph always being inside the bold blue graph).

**Table 5** The evaluation results of the MTR criterion, which compares CDD methods with methods based on ER monitor using the Friedman test

| methodology / dataset | raw_base | IME_base | BD_base | rIME_base | rBD_base |
|---|---|---|---|---|---|
| accumulo | 1500 | 1500 | 1500 | 6000 | 6000 |
| angular | 221 | 239 | 221 | 239 | 221 |
| brackets | 279 | 139 | 279 | 11700 | 11700 |
| bugzilla | 4800 | 4800 | 4800 | 3200 | 3200 |
| camel | 2193 | 1919 | 2193 | 10233 | 10233 |
| cinder | 370 | 370 | 370 | 7400 | 7400 |
| django | 6300 | 4200 | 1232 | 4200 | 1938 |
| gephi | 643 | 643 | 643 | 643 | 643 |
| hibernate-orm | 4050 | 2700 | 4050 | 579 | 225 |
| hibernate-search | 750 | 750 | 750 | 857 | 857 |
| imglib2 | 575 | 575 | 575 | 177 | 177 |
| jetty | 620 | 620 | 620 | 12400 | 1771 |
| kylin | 6000 | 6000 | 6000 | 6000 | 500 |
| log4j | 1000 | 1000 | 1000 | 1000 | 1000 |
| nova | 2690 | 6725 | 6725 | 13450 | 13450 |
| postgres | 732 | 2090 | 1819 | 4390 | 2665 |
| tomcat | 2350 | 2089 | 2089 | 895 | 1567 |
| wordpress | 4688 | 9375 | 4312 | 6250 | 4688 |
| meanAvg | 4.22 | 4.11 | 4.5 | 3.17 | 3.94 |

**Table 6** The evaluation results of the MTR criterion, which compares CDD methods with methods based on various criteria monitor using Friedman test

| | raw_base | IME_base | BD_base | rIME_base | rBD_base |
|---|---|---|---|---|---|
| ER PH-Er PH | 4.22 | 4.11 | 4.5 | 3.17 | 3.94 |
| AUC-Er PH | 4.33 | 3.75 | 4.67 | 3.42 | 3.61 |
| Gmean-Er PH | 4.06 | 3.17 | 4.06 | 3.83 | 4.75 |
| Precision-Er PH | 4.08 | 3.56 | 4.81 | 3.61 | 4.08 |
| Recall-Er PH | 4.28 | 3.28 | 3.89 | 3.89 | 4.61 |
| MCC-Er PH | 4.31 | 3.31 | 4.25 | 3.14 | 4.03 |
| Fmeasure-Er PH | 4.22 | 3.03 | 4.31 | 3.81 | 4.33 |

**Table 7** The evaluation results of the cdAccuracy criterion, which compares CDD methods with methods based on various criteria monitor using Friedman test (no significant difference compared to MTR and Raw_base performed better)

|              | raw_base | IME_base | BD_base | rIME_base | rBD_base |
|--------------|----------|----------|---------|-----------|----------|
| ER PH-Er PH  | 3.69     | 3.89     | 4.33    | 3.89      | 4.22     |
| AUC-Er PH    | 3.75     | 3.94     | 4.42    | 3.86      | 3.94     |
| Gmean-Er PH  | 3.75     | 3.75     | 4.44    | 3.86      | 4.17     |
| Precision-Er PH | 3.75  | 3.86     | 4.44    | 3.75      | 4.14     |
| Recall-Er PH | 3.69     | 3.69     | 4.36    | 3.89      | 4.31     |
| MCC-Er PH    | 3.72     | 3.92     | 4.42    | 3.64      | 3.94     |
| Fmeasure-Er PH | 3.75   | 3.75     | 4.44    | 3.86      | 4.19     |

**Table 8** The evaluation results of the MTD criterion, which compares CDD methods with methods based on various criteria monitor using Friedman test (no significant difference compared to MTR)

|              | raw_base | IME_base | BD_base | rIME_base | rBD_base |
|--------------|----------|----------|---------|-----------|----------|
| ER PH        | 4.03     | 4.42     | 4.31    | 3.28      | 3.72     |
| AUC-Er PH    | 4.19     | 3.81     | 4.58    | 3.31      | 3.94     |
| Gmean-Er PH  | 3.89     | 3.5      | 3.94    | 4.03      | 4.56     |
| Precision-Er PH | 3.79  | 3.71     | 4.44    | 3.76      | 4.5      |
| Recall-Er PH | 4.03     | 3.64     | 3.89    | 3.83      | 4.5      |
| MCC-Er PH    | 4.17     | 3.69     | 4.17    | 3.19      | 3.92     |
| Fmeasure-Er PH | 3.91   | 3.29     | 4.18    | 3.79      | 4.59     |

**Table 9** The evaluation results of the MTFA criterion, which compares CDD methods with methods based on various criteria monitor using Friedman test (BD_base performed better than IME_base in rebalanced ones).

|              | raw_base | IME_base | BD_base | rIME_base | rBD_base |
|--------------|----------|----------|---------|-----------|----------|
| ER PH        | 4.64     | 4.08     | 4.36    | 3.33      | 3.28     |
| AUC-Er PH    | 4.53     | 3.94     | 4.33    | 3.75      | 3.36     |
| Gmean-Er PH  | 4.36     | 3.58     | 4.17    | 4.17      | 3.97     |
| Precision-Er PH | 4.35  | 3.74     | 4.56    | 3.74      | 3.74     |
| Recall-Er PH | 4.33     | 3.56     | 4.14    | 3.94      | 4.14     |
| MCC-Er PH    | 4.33     | 3.75     | 4.14    | 3.94      | 3.75     |
| Fmeasure-Er PH | 4.5    | 3.72     | 4.31    | 3.72      | 3.72     |

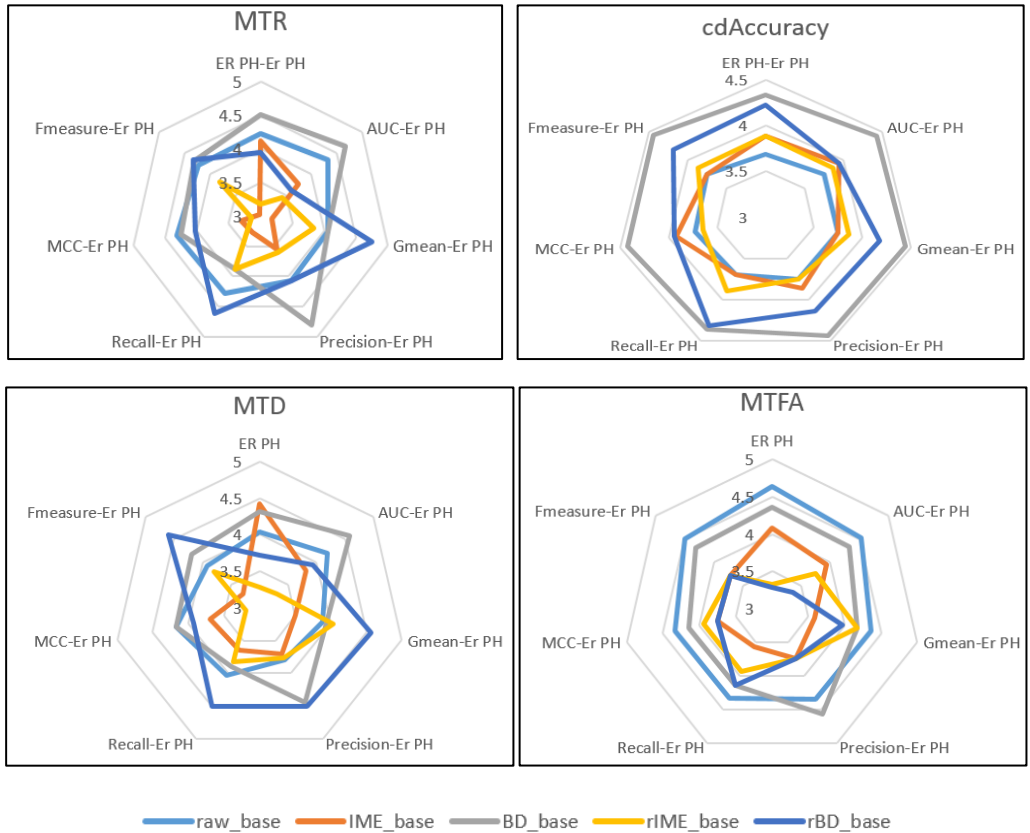

**Fig. 5** the radar diagram related to Tables 6 to 9

> RQ1: Given that one of the key advantages of our proposed method is its ability to locate CD on unlabeled data, we have developed two methods based on instance interpretation. As a result, the method based on IME interpretation outperformed the BD_based method. Furthermore, even in interpretation-based method on rebalanced data, IME_based method still demonstrated superior performance.

## 4.2 Exploring CDD methods based on monitoring criteria other than ER and their impact on the proposed methods based on data resampling (RQ2)

As previously mentioned, detecting of CD points using ER monitoring is common. However, in this work, we have discovered CD points using monitoring of metrics other than ER. Additionally, to more extensively examine the performance of the proposed methods based on interpretation, we have evaluated their compatibility with the baseline methods associated with monitoring the performance of different criteria using four well-known performance criteria of CDD methods; MTR, cdAccuracy, MTD, and MTFA. As stated earlier, the criteria used in the baseline methods are either threshold-dependent or threshold-independent. In the previous section, we noted that studies have shown that the performance of each of these two categories on the stability of online models differs. Therefore, relevant studies avoid them. On the other hand, checking model stability on resampled data is also a challenge in this field. To address these two challenges, we posed RQ2. To address this question, we investigated the consistency of CD points obtained from baseline methods with those obtained from proposed methods implemented on simplified and rebalanced data. The results are presented in Table 6 to Table 9. To provide a clearer response, we have categorized the results of these tables into two groups: threshold-dependent and threshold-independent criterion-based methods using radar charts. These charts are displayed in Figure 6 (related to the method based on threshold-independent criteria) and Figure 7 (related to the method based on threshold-dependent criteria). In view of Figure 6, when using threshold-independent criteria for baseline methods, the output of proposed methods implemented on resampling data is more consistent with the results of these baseline methods according to all four famous criteria of CDD methods: MTR, cdAccuracy, MTD, and MTFA. As shown in this figure, the yellow chart (rIME-base) consistency falls within the orange chart (IME-base), while the bold blue chart (rBD-base) consistency falls within the gray chart (BD-base).

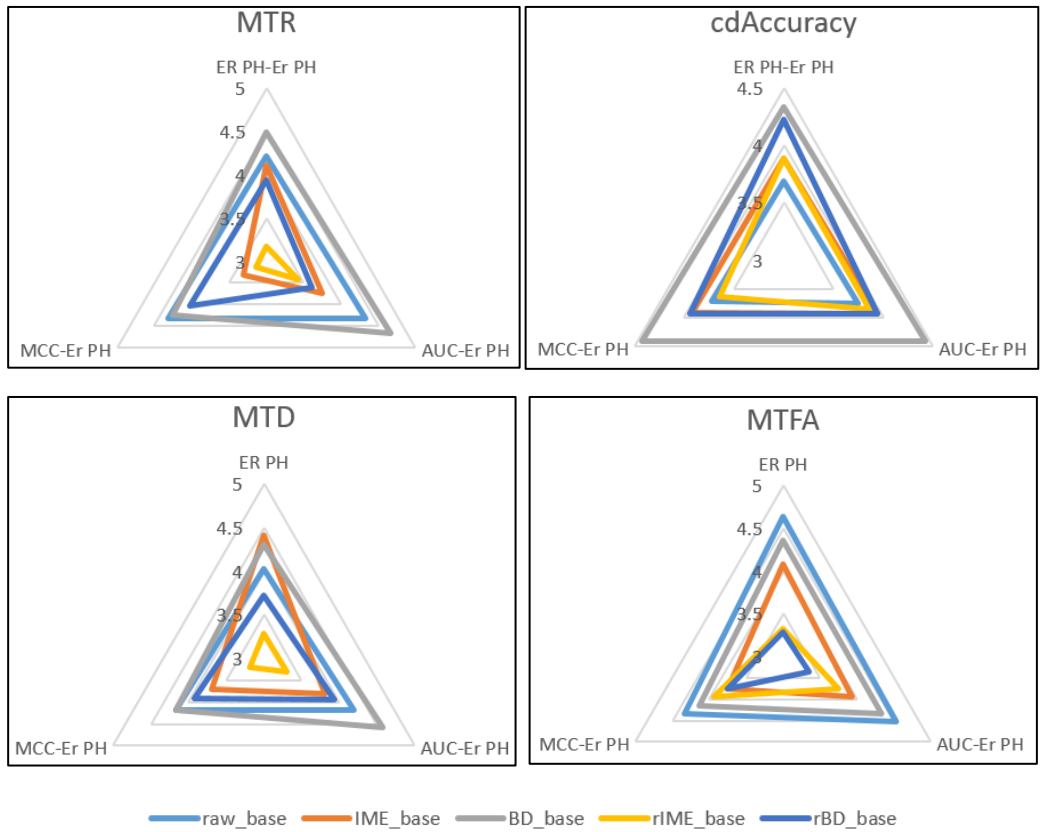

**Fig. 6** Comparison of the compatibility between proposed methods and methods based on threshold-independent criteria using four efficiency criteria of the CDD method

According to Figure 7, when using threshold-dependent criteria for the baseline methods, the output of the proposed methods implemented on simple data is more consistent with the results of these baseline methods than on the resampled data. The orange graph (IME-base) always falls within the yellow graph (rIME-base) as shown in this figure. However, the accuracy of the CDD method based on BD_base interpretation is lower than rBD_base, which affects the false alarm criterion, i.e., MTFA.

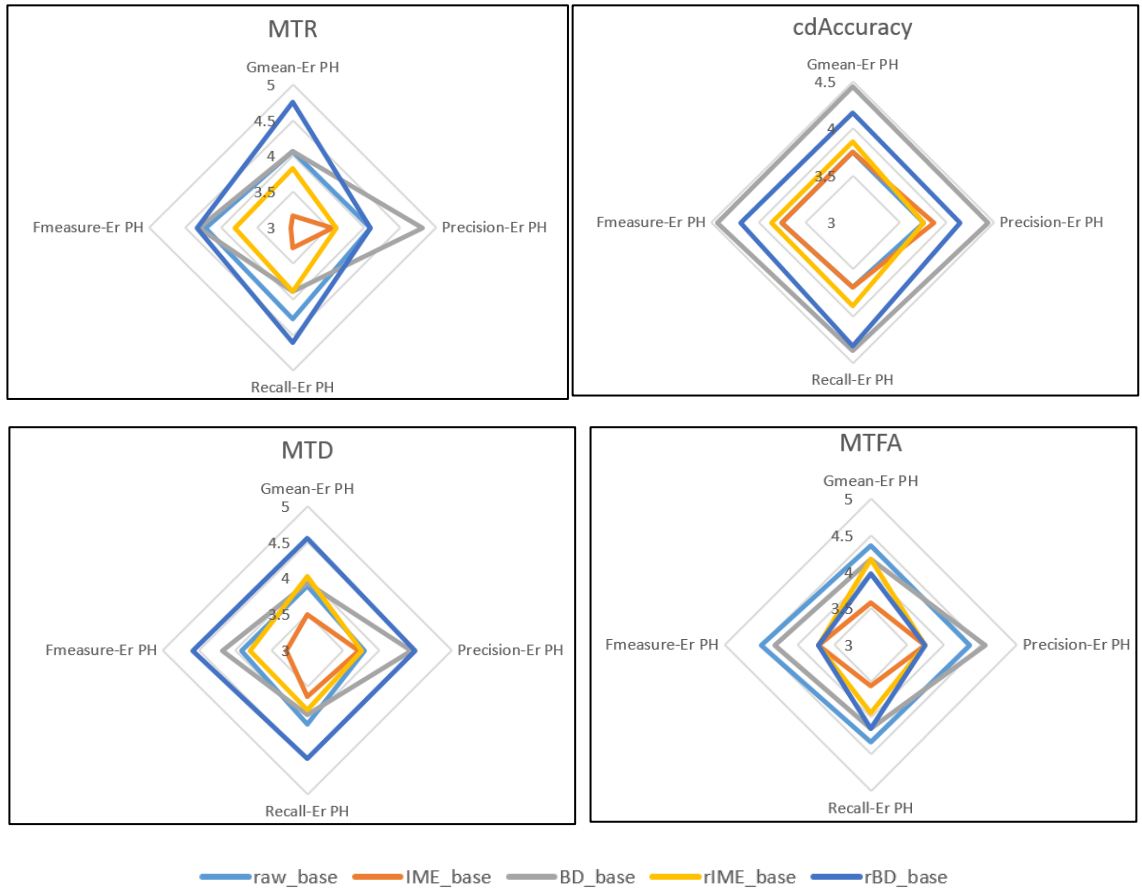

**Fig. 7** Comparison of the compatibility between proposed methods and methods based on threshold-dependent criteria using four efficiency criteria of the CDD method

Table 10 The performance of the Random Forest classifier before and after resampling

|  | Before | After |
|---|---|---|
| Accuracy | 12 | 3 |
| AUC | 13 | 3 |
| Specificity | 15 | 4 |
| AverageCost | 3 | 12 |
| BrierScore | 2 | 12 |
| InformationScore | 4 | 16 |
| Sensitivity | 4 | 14 |
| Kappa | 8 | 6 |
| Precision | 15 | 1 |
| Recall | 4 | 14 |
| Fmeasure | 4 | 10 |
| Gmean | 5 | 11 |
| KS | 10 | 2 |
| TPR | 2 | 4 |
| FPR | 0 | 6 |

**Examining the classification performance before and after resampling:**

Table 10 presents the number of times the performance of the Random Forest classifier was better before and after resampling. For instance, the first row shows that out of 20 datasets, the accuracy measure for 12 datasets displayed better performance before resampling, while the accuracy of 3 datasets improved after resampling. These results indicate that resampling reduces the value of threshold-independent performance criteria, while enhances the value of threshold-dependent criteria after resampling. We conducted this analysis using the Wilcoxon statistical test.

> RQ2: Threshold-independent performance measures of the Random Forest classifier decrease after resampling, while its threshold-dependent criteria improve. Resampling also reduces the matching of the CD points discovered by proposed CDD methods with the output of baseline methods based on monitoring threshold-dependent criteria. However, it improves compared to baseline methods based on monitoring threshold-independent measures.

## 4.3 Comparison of the method based on repeated predictions with and without the class labels to interpretation-based methods (RQ3)

Our other proposed method compares the multivariate one-way ANOVA of repeated predictions. Table 11 displays the output of repeated predictions on each test sample. The first column represents the sample number, the second column shows the label of each sample, and the third to twelfth columns display the probability of prediction (PP stands for Prediction_Probability). The last column indicates the group to which the test sample belongs.

Table 11 Partial representation of the repeated predictions output for one of the datasets

| Instance Number | Label | PP1 | PP2 | PP3 | PP4 | PP5 | PP6 | PP7 | PP8 | PP9 | PP10 | Group |
|---|---|---|---|---|---|---|---|---|---|---|---|---|
| 1 | 1 | 0.41 | 0.5 | 0.43 | 0.53 | 0.5 | 0.49 | 0.53 | 0.52 | 0.48 | 0.42 | 0 |
| 2 | 1 | 0.91 | 0.91 | 0.91 | 0.91 | 0.9 | 0.93 | 0.9 | 0.9 | 0.91 | 0.92 | 0 |
| 3 | 1 | 0.76 | 0.78 | 0.8 | 0.73 | 0.71 | 0.7 | 0.78 | 0.75 | 0.78 | 0.75 | 0 |
| . | . | | | | . | | | | | | | . |
| . | . | | | | . | | | | | | | . |
| . | . | | | | . | | | | | | | . |
| 98 | 1 | 0.8 | 0.8 | 0.8 | 0.82 | 0.78 | 0.78 | 0.81 | 0.83 | 0.81 | 0.81 | 0 |
| 99 | 0 | 0.3 | 0.28 | 0.28 | 0.27 | 0.32 | 0.31 | 0.25 | 0.27 | 0.28 | 0.26 | 0 |
| 100 | 1 | 0.93 | 0.95 | 0.93 | 0.92 | 0.92 | 0.95 | 0.94 | 0.93 | 0.94 | 0.93 | 0 |
| 101 | 1 | 0.94 | 0.94 | 0.93 | 0.91 | 0.94 | 0.93 | 0.93 | 0.93 | 0.94 | 0.92 | 1 |
| 102 | 1 | 0.57 | 0.52 | 0.63 | 0.56 | 0.59 | 0.57 | 0.59 | 0.57 | 0.61 | 0.57 | 1 |
| . | . | | | | . | | | | | | | 1 |
| . | . | | | | . | | | | | | | 1 |
| . | . | | | | . | | | | | | | 1 |

Table 12 illustrates a significant distribution difference of repeated predictions obtained through the multivariate one-way ANOVA method between Group 10 (the target group) and six out of nine preceding groups (as indicated by a p-value < 0.05).

**Table 12** A demonstration of the statistical distribution differences of a group compared to its previous groups

| Target Group | Primary Groups | P-value |
|---|---|---|
| 10.0 | .0 | 0.443 |
| | 1.0 | 0.000 |
| | 2.0 | 0.000 |
| | 3.0 | 0.009 |
| | 4.0 | 0.000 |
| | 5.0 | 0.000 |
| | 6.0 | 0.000 |
| | 7.0 | 0.311 |
| | 8.0 | 0.096 |
| | 9.0 | 0.151 |

Table 13 indicates that each group has a significant difference in terms of statistical distribution with how many of its previous groups. For instance, group ten displays a statistical difference from six of its preceding groups. However, as this difference is one for the groups before and after it, it does not qualify as a CD point and is considered a random event by the PH algorithm. Consequently, updating the model at this point is unnecessary. Table 12 and Table 13 present the output related to the "Accumula" project.

**Table 13** The number of groups that have a statistical difference with each group

| Group Num | 1 | 2 | 3 | 4 | 5 | 6 | 7 | 8 | 9 | 10 | 11 | 12 | 13 | 14 | 15 | 16 | 17 | 18 | 19 | 20 |
|---|---|---|---|---|---|---|---|---|---|---|---|---|---|---|---|---|---|---|---|---|
| # Dis Diff | 0 | 0 | 0 | 1 | 1 | 1 | 1 | 1 | 1 | 6 | 4 | 1 | 1 | 3 | 0 | 0 | 0 | 1 | 0 | 1 |

Table 14 compares the accuracy of the method based on repeated predictions in four modes, with and without the class label, and using simple and resampled data, with the baseline methods using the Friedman statistical test. This proposed method is referred to as Pred for short, with the letter "c" at the end indicating it is based on labeled data, and the letter "R" at the beginning indicating resampled input data. The ranking results of the proposed methods based on other CDD evaluation criteria using the Friedman test are presented in Table 15 to Table 17. In this section, we will provide a summary and analysis of the proposed method based on repeated predictions. We will compare the performance of the proposed method with and without class labels on both simple and resampled data, as shown in Table 18 and Table **19**. Similar to the interpretation-based method, the results obtained from the baseline threshold-dependent methods show that the proposed method performs better on simple data (pred_c, pred) than it does on resampled data (rpred_c, rpred). Although Pred_c generally outperforms Pred, their results do not differ significantly. Therefore, the method based on without the class label is also suitable.

**Table 14** A comparison of CDD_Accuracy measure of different methods based on repeated predictions using Friedman test

|  | Pred_c | Pred | Rpred_c | Rpred |
| --- | --- | --- | --- | --- |
| ER PH | 3.22 | 3.56 | 3.58 | 3.97 |
| AUC-Er PH | 3.22 | 3.39 | 3.36 | 3.86 |
| Gmean-Er PH | 3.11 | 3.56 | 3.39 | 3.53 |
| Precision-Er PH | 3.14 | 3.47 | 3.47 | 3.81 |
| Recall-Er PH | 3.08 | 3.53 | 3.36 | 3.67 |
| MCC-Er PH | 3.22 | 3.5 | 3.17 | 3.64 |
| Fmeasure-Er PH | 3 | 3.47 | 3.47 | 3.81 |

**Table 15** A comparison of MTD measure of different methods based on repeated predictions using Friedman test

|  | Pred_c | Pred | Rpred_c | Rpred |
| --- | --- | --- | --- | --- |
| ER PH | 3.56 | 3.75 | 3.17 | 3.19 |
| AUC-Er PH | 3.19 | 3.53 | 3.31 | 3.42 |
| Gmean-Er PH | 2.81 | 3.33 | 3.78 | 3.67 |
| Precision-Er PH | 3.32 | 3.21 | 3.47 | 3.47 |
| Recall-Er PH | 2.94 | 3.31 | 4.06 | 3.56 |
| MCC-Er PH | 3.47 | 3.33 | 3.39 | 3.19 |
| Fmeasure-Er PH | 3.12 | 3.15 | 3.76 | 3.62 |

**Table 16** A comparison of MTFA measure of different methods based on repeated predictions using Friedman test

|  | Pred_c | Pred | Rpred_c | Rpred |
| --- | --- | --- | --- | --- |
| ER PH | 3.5 | 3.67 | 3.11 | 3.44 |
| AUC-Er PH | 3.42 | 3.42 | 3.25 | 3.58 |
| Gmean-Er PH | 3.14 | 3.42 | 3.58 | 3.56 |
| Precision-Er PH | 3.22 | 3.39 | 3.56 | 3.72 |
| Recall-Er PH | 3.17 | 3.44 | 3.44 | 3.58 |
| MCC-Er PH | 3.25 | 3.36 | 3.36 | 3.67 |
| Fmeasure-Er PH | 3.19 | 3.5 | 3.36 | 3.53 |

**Table 17** A comparison MTR measure of different methods based on repeated predictions using Friedman test

|  | Pred_c | Pred | Rpred_c | Rpred |
|---|---|---|---|---|
| ER PH | 3.31 | 3.75 | 3.44 | 3.53 |
| AUC-Er PH | 3.14 | 3.5 | 3.28 | 3.53 |
| Gmean-Er PH | 2.75 | 3.33 | 3.97 | 3.72 |
| Precision-Er PH | 3.22 | 3.22 | 3.61 | 3.78 |
| Recall-Er PH | 2.97 | 3.39 | 3.92 | 83.56 |
| MCC-Er PH | 3.47 | 3.36 | 3.22 | 3.33 |
| Fmeasure-Er PH | 3.12 | 3.38 | 3.44 | 3.65 |

**Table 18** A comparison between rpred_c -based and pred_c-based methods

|  | CDD_Accuracy | MTD | MTFA | MTR |
|---|---|---|---|---|
| pred_c better | All except MCC | All except ER, MCC | All except ER, AUC | All except MCC |
| pred_c = rpred_c | - | - | - | - |
| rpred_c better | MCC | ER, MCC | ER, AUC | MCC |

**Table 19** A comparison between rpred -based and pred-based methods

|  | CDD_Accuracy | MTD | MTFA | MTR |
|---|---|---|---|---|
| pred better | All except Gmean | All except ER, AUC, MCC | All except ER | All except ER, MCC |
| pred = rpred | Gmean | - | - | - |
| rpred better | - | ER, AUC, MCC | ER | ER, MCC |

**Comparing all proposed methods with the method based on model interpretation:**

In this section, we compare the proposed methods based on instance interpretation with the method based on model interpretation proposed by the study of (Demšar and Bosnić 2018). The IME algorithm can extract three types of model interpretation vectors: the positive effect size of features on the target class, the negative effect size of features, and the average effect size of features. Demšar and Bosnić used the last case to detect CDs, but our previous work (Chitsazian, Sedighian Kashi et al. 2023), has shown that using the first two cases is more effective. When comparing CDD_Accuracy, our proposed methods based on instance interpretation and Pred outperform all three types of model interpretation. Among these three types, the positive effect size consistently performs best while the negative effect size consistently performs worst. The positive effect size is better than BD only in Gmean-Er PH, Recall-Er PH, and F-Measure-Er PH. In terms of MTD performance, all types of model interpretation are last with the positive effect size being consistently best among them. In MTFA performance, these three types are worst but the average effect size is best among them. Finally, in MTR performance, these three types have worst performance overall.

> RQ3: After evaluating the MTFA and MTR measures, it was discovered that the performance of 'Pred' and 'Pred_c' is superior to IME interpretation. Moreover, in all four performance measures, both 'Pred' and 'Pred_c' outperform BD interpretation. Additionally, all three model interpretation types perform worse than our proposed methods, which are based on instance interpretation and Preds.

# 5 Treats to validity

In any experimental study, threats to the validity of the concluded results and its evaluations should be analyzed. The following are the threats of the validity of this research, which include three categories: construct, internal, and external.

## 5.1 Construct Validity

Threats to construct validity are related to the alignment of our chosen indicators with what we intend to measure. In our method, we have employed a widely-used rebalancing algorithm called SMOTE in our method and assessed the approach using various performance metrics and two interpretation methods. One of the studies mentioned demonstrated that the rebalancing techniques consistently work with different parameters. As our objective is not to explore different rebalancing techniques, there is no need to investigate other techniques. Our analysis of the Friedman statistical test algorithm revealed a high false alarm rate for default values and values smaller than that. This prompted us to explore larger parameter values for optimal results. For other domains with higher data entry rates, different parameter values may be useful. However, since we conducted our experiments on 20 datasets of JIT-SDP and evaluated them with statistical tests on different efficiency criteria, we can conclude that it is efficient for datasets of this size in the JIT_SDP problem. Nevertheless, it can be separately examined on larger datasets in this area because automatic parameter optimization techniques are costly. We conducted our analysis using Random Forest models, but regression models and other classification techniques such as Naive Bayes can also be examined.

## 5.2 Internal validity

In terms of internal validity, we demonstrate the extent to which our hypotheses and methods can account for the results obtained. For this study, we utilized a large dataset that has been employed in various valid studies to investigate and validate our findings. One of our objectives is to identify concept drift using unlabeled input data and compare it with methods based on labeled data. However, since other fields have used the ER criterion-based method to detect concept drift, we also compared our approach with this baseline method and used the error rates from other criteria instead of ER to support our results. Specifically, we found that our approach and the ER criterion-based method both yielded similar outcomes in detecting concept drift. As previously mentioned, a study on the discovery of CD points revealed that interpretation-based methods outperform those based on model performance. In this work, we also employed the interpretation algorithm used in their method and compared it with the interpretation method presented in a JIT-SDP article as a compatible interpretation algorithm. Both methods produced identical outputs, indicating their compatibility and effectiveness in identifying CD points.

## 5.3 External validity

External validity refers to the ability to generalize proposed methods to other studies and regions. Given that our research focuses on an online fault diagnosis system (FDS), it can be extended to other types of FDS systems, such as chemical fault diagnosis (Wu and Zhao 2018) and diagnosing defects in induction motors (Razavi-Far, Farajzadeh-Zanjani et al. 2017), among others. Our study was conducted on 20 commonly used datasets in JIT-SDP literature, all of which share the same metrics. Therefore, it cannot be extrapolated to datasets with more diverse criteria.

# 6 Conclusion and future works

In this paper, we present the results of the first empirical study on discovering concept drift points in simple and resampled JIT-SDP data, both with and without labels. Our study reveals that the performance of JIT-SDP models is unstable over time. Previous research on SDP stability and CDD has focused on labeled input data. However, obtaining newly incoming labeled data can be difficult in real-world scenarios. To address this issue, we propose new unsupervised CDD methods for JIT-SDP input data and evaluate them using 20 datasets. Another study on stream data has shown that an interpretation-based approach outperforms distribution-based methods. Therefore, we adopt this approach for CDD in the JIT-SDP domain and demonstrate significant improvements, including handling unlabeled data and comparing it with commonly proposed CDD methods in literature. Additionally, due to class imbalance issues in SDP problems, class rebalancing has always been a concern for researchers in this field. For this reason, we also investigate the impact of data rebalancing on CDD by comparing it with methods based on threshold-independent and threshold-dependent metrics.

When it is important to improve the performance of threshold-dependent metrics and increase the completeness of detecting software defects, such as Recall (Tantithamthavorn, Hassan et al. 2018), methods based on simple and unbalanced data outperform, and the output of the proposed method is closer to the baseline method. Since we derived this result using several threshold-dependent metrics, it can be concluded that it is not random. However, if performance metrics independent of threshold such as ER and AUC, which are considered standard metrics for SDP models in many studies, are important, rebalancing the data leads to better model performance. Relevant research has used model interpretation to discover CD points, which requires labeling of input data. In contrast, we have used methods based on instance interpretation that perform better than model interpretation and do not require labeling of input data to discover CD points. Moreover, we conclude that extracting the instance interpretation vector using the IME algorithm works better than the BD algorithm for discovering CD. Another CDD proposed method based on repeated prediction distribution analysis over time performs well on both labeled and unlabeled data and outperforms other methods as mentioned in the results section.

This research has been conducted on within-project JIT-SDP datasets. In future works, the proposed methods can be investigated using integrated datasets from different projects. In related works, the model is trained using the latest input data before the testing data after concept drift occurs. To improve results, we can extract relevant data suitable for training the online classifier after detecting CD. In this study, we used a simple classifier in the CDD method; however, in future works, more complex classifications such as deep learning or multi-objective optimization algorithms can be used to improve results. Additionally, the effectiveness of the proposed methods in this research can also be investigated on streaming data from non-software error detection system.